\RequirePackage{ifpdf} 
\ifpdf
\documentclass[pdftex,12pt,a4paper]{article}  
\else
\documentclass[12pt,a4paper]{article} 
\fi

\parskip=\medskipamount

\usepackage[mathscr]{euscript} 	
\usepackage{bbm} 
\usepackage{amsfonts,amsmath,amssymb}
\ifpdf
\usepackage[pdftex]{graphicx}	
\usepackage[usenames,pdftex]{color}		
\usepackage[pdftex,bookmarks]{hyperref}
\hypersetup{
		bookmarksopen,bookmarksnumbered,
		pdfnewwindow=true,
		pdfauthor = {Simon J. Tyler and Sergei M. Kuzenko},
		pdftitle = {Relating the Komargodski-Seiberg and Akulov-Volkov actions},
		pdfsubject = {Supersymmetry},
		pdfkeywords = {supersymmetry, Goldstino, Akulov-Volkov},
		pdfcreator = {LaTeX with hyperref package},
		pdfproducer = {pdflatex}
}
\else
\usepackage[dvips]{graphicx}
\usepackage[usenames,dvips]{color}
\usepackage[dvips,bookmarks,colorlinks=false]{hyperref}
\fi
\def\a{\alpha}
\def\b{\beta}
\def\c{\chi}
\def\d{\delta}

\def\g{\gamma}

\def\j{\psi}
\def\k{\kappa}
\def\l{\lambda}

\def\s{\sigma}

\def\x{\xi}
\def\z{\zeta}

\def\X{\Xi}

\newcommand{\db}{{\dot{\beta}}}

\def\ds1{\ensuremath{\mathbbm{1}}}
\newcommand{\rmd}{{\rm d}}

\newcommand{\rmi}{{\rm i}}
\newcommand{\rmr}{{\rm r}}

\newcommand{\expt}[1]{\ensuremath{\left< #1 \right>}}

\renewcommand{\Re}{\ensuremath{\text{Re}}}

\newcommand{\pd}{\partial}
 				
\def\tr{{\rm tr}}		

\def\cc {{\rm c.c.}}

\def\intx{\int\!\!{\rmd}^4x\,}

\def\half{\ensuremath{\frac{1}{2}}} 

\newcommand{\be}{\begin{equation}}
\newcommand{\ee}{\end{equation}}
\newcommand{\bea}{\begin{eqnarray}}
\newcommand{\eea}{\end{eqnarray}}
\newcommand{\non}{\nonumber}
\newcommand{\bm}[1]{\mbox{\boldmath$#1$}}
\newcommand{\AV}{{\mathrm{AV}}}
\newcommand{\KS}{{\mathrm{KS}}}

\newcommand{\vb}{{\bar{v}}}
\newcommand{\ub}{{\bar{u}}}
\author{Simon Tyler and Sergei Kuzenko} 
\date{September 2010}                   
\begin{document}                        
\begin{titlepage}

\begin{flushright}
September, 2010
\end{flushright}


\begin{center}
{\large \bf  Relating the Komargodski-Seiberg and Akulov-Volkov actions: 
Exact nonlinear field redefinition}
\end{center}
\begin{center}
{\large  
{Sergei M. Kuzenko}
\footnote{kuzenko@cyllene.uwa.edu.au}
and 
{Simon J.\ Tyler}
\footnote{styler@physics.uwa.edu.au}

\vspace{5mm}

\footnotesize{
{\it School of Physics M013, The University of Western Australia\\
35 Stirling Highway, Crawley W.A. 6009, Australia}}  

\vspace{2mm}}
\end{center}

\vspace{5mm}

\pdfbookmark[1]{Abstract}{abstract_bookmark}
\begin{abstract}
\baselineskip=14pt
This paper constructs an exact field redefinition that maps the 
Akulov-Volkov action to that recently studied by  
Komargodski and Seiberg in arXiv:0907.2441.
It is also shown that the approach advocated in 
arXiv:1003.4143v2 and arXiv:1009.2166
for deriving such a relationship is inconsistent.
\end{abstract}
\vfill
\end{titlepage}

\section{Introduction}
The Akulov-Volkov (AV) action \cite{VA} is the second oldest 
supersymmetric theory in four space-time dimensions.
It describes the low-energy dynamics of a massless Nambu-Goldstone fermionic
particle which is associated with the spontaneous breaking of rigid 
supersymmetry and  is called the Goldstino 
(see \cite{WB} for a nice review of the AV model and related concepts).
According to the general theory of the nonlinear realization of 
${\cal N}=1$ supersymmetry 
\cite{VA,Ivanov1977,Ivanov1978,Ivanov1982,Uematsu:1981rj}, 
the AV action is universal in the sense that 
any Goldstino model should be related to the AV action by a 
nonlinear field redefinition.
Various Goldstino models can be  interesting in their own right,\footnote%
{
For instance, the fermionic sector of the ${\cal N}=1$  
supersymmetric Born-Infeld action \cite{CF,BG} is a new Goldstino model. 
It has been shown to be related to the AV action by 
a nonlinear field redefinition \cite{HK,Kuzenko2005e}.
}
in particular, 
those models which are 
realized in terms of  constrained superfields.
Over the years, there have appeared a number of such  superfield actions with 
spontaneously broken supersymmetry
\cite{Rocek1978, LR,Samuel1983,Casalbuoni1989}. 

Recently there has been renewed interest in Goldstino couplings inspired by  
the work of Komargodski and Seiberg \cite{Komargodski2009}. 
They put forward the Goldstino model 
that had actually appeared in the literature twenty years earlier 
\cite{Casalbuoni1989}. 
The novelty of the Komargodski-Seiberg (KS) approach is that they related the 
Goldstino dynamics to the superconformal anomaly multiplet 
$X$ corresponding to the 
Ferrara-Zumino supercurrent \cite{FZ}.
Under the renormalization group flow, the multiplet of anomalies 
$X$ defined in the
UV turns out to flow in the IR to a  chiral superfield $X_{NL}$ 
(obeying the constraint $X_{NL}^2=0$, 
of the type first introduced by Ro\v{c}ek \cite{Rocek1978}) 
which contains the Goldstino as  a component field. For a ${\cal N}=2$ 
generalization of the KS formalism, see \cite{AB}.

The action derived in \cite{Casalbuoni1989,Komargodski2009}
has a particularly simple form both in superspace
and when reduced to components.  
However, its direct relation to the AV action and thus the structure of its 
nonlinearly realized  supersymmetry have not yet been studied.

In \cite{Liu2010} it was shown, using the general method of  
\cite{Ivanov1978, Ivanov1982}, 
that the Goldstino action introduced by Samuel and Wess \cite{Samuel1983}
can be derived from the constrained superfield formalism of 
Komargodski and Seiberg. The former model is known to be equivalent to 
the AV theory \cite{Samuel1983}.

What we provide in this paper is a direct relation between the AV
action and that of Komargodski-Seiberg. 
Unlike  \cite{Liu2010}, we do not make use of the techniques  developed in  
\cite{Ivanov1978, Ivanov1982}. 
Instead we follow the approach  pursued in \cite{Kuzenko2005e} 
which can also be applied to study the fermionic sector of 
supersymmetric Euler-Heisenberg-type actions.
We use the two-component notation and conventions adopted in \cite{WB,BK}.

\pagebreak[3]
\section{Setup and results}
The AV action \cite{VA} is
\begin{align}
\label{defn:AV1}
	S_\AV[\l,\bar\l] &= \frac1{2\k^2}\intx\Big(1-\det\X\Big)~,
\end{align}
where $\k$ denotes  the dimensionful coupling constant and
\begin{align}
	\X_a{}^b 
			&= \d_a{}^b+{\k^2}\left(v+\bar{v}\right)_a{}^b~,&
	v_a{}^b	&=\rmi\l\s^b\pd_a\bar\l ~, \quad  \quad 
	\vb_a{}^b	=-\rmi\pd_a\l\s^b\bar\l~.
\end{align}
By construction, $S_\AV $ is invariant under the 
nonlinear supersymmetry transformations
\begin{align} \label{eqn:Nonlin_SUSY}
		\d_\x\l_\a &= \frac1\k\x_\a 
			-\rmi\k \big(\l\s^a\bar\x-\x\s^a\bar\l\big)\pd_a\l_\a ~.
\end{align}
Expanding out the determinant in \eqref{defn:AV1}
and denoting the trace of a  matrix $M=(M_a{}^b)$ with Lorentz indices as
\( \expt{M} = \tr(M) = M_a{}^a \)
yields
\begin{align} \label{defn:AV2}
	S_\AV[\l,\bar\l] = -\half&\intx\Bigg(\expt{v+\bar v}
	+2\k^2\Big(\expt{v}\expt{\bar v}-\expt{v\bar v}\Big)\non \\
	&+{\k^4}\Big(\expt{v^2\vb}-\expt{v}\expt{v\vb}
		-\half\expt{v^2}\expt{\vb}+\half\expt{v}^2\expt{\vb}+\cc\Big)\Bigg)~.
\end{align}
As demonstrated in \cite{Kuzenko2005e}, the 8th-order terms vanish.

The Goldstino action constructed in 
\cite{Casalbuoni1989,Komargodski2009}
is
\begin{equation}\label{defn:KS1}
	S_{\KS}[\j,\bar\j] = -\half\intx\Big(\expt{u+\bar u}+
		\frac1{2f^2}\pd^a\bar{\j}^2\pd_a \j^2
		+\frac1{8f^6}\j^2\bar\j^2\pd^2\j^2\pd^2\bar\j^2\Big)~,
\end{equation}
where 
we defined $u_a{}^b = \rmi\j\s^b\pd_a\bar\j$
and its complex conjugate. In the following section, we find that
the constant $f$ is related to $\k$ via $2f^2=\k^{-2}$.

Below, we find that the nonlinear field redefinition
which maps the action \eqref{defn:AV1} to the action \eqref{defn:KS1}, i.e.\
$S_\AV[\l_\a (\j,\bar\j),\l_\a (\j,\bar\j)] = S_\KS[\j,\bar\j]$,
can be chosen to be
\begin{align} \label{eqn:AVtoNL_simp}
	\l_\a (\j,\bar\j) 
	&= \j_\a - \rmi\frac{\k^2}2(\s^a\bar\j)_\a(\pd_a\j^2) 
	- \frac{\k^4}{2}\j_\a\Big(\expt{u\ub}-2\expt{u}\expt{\ub}
		+\frac12\expt{\ub^2} - \frac12\pd^a\j^2\pd_a\bar\j^2  \non\\
	&\qquad +\frac14\bar\j^2\Box\j^2\Big) 
	+ {\k^6}{}\j_\a\big(\expt{u\ub^2}+\frac3{2}\expt{u \ub}\expt{\ub}
		+ \frac3{4}\expt{u}\expt{\ub^2}  \big) ~.
\end{align}
The inverse field redefinition is 
\begin{align} \label{eqn:NLtoAV_simp}
	\j_\a (\l,\bar\l) 
	&= \l_\a + \rmi\frac{\k^2}2(\s^a\bar\l)_\a(\pd_a\l^2)
	\Big(1 + {\k^2}\expt{\vb} \Big) 				\non \\
	&+ \frac{\k^4}{2}\l_\a\big(\expt{v\vb}
		-\frac12\expt{\vb^2} - \expt{\vb}^2
		+ \frac12\pd^a\l^2\pd_a\bar\l^2 + \frac34\bar\l^2\Box\l^2\big) \\\non
	&- {\k^6}{}\l_\a\big(\expt{v\vb^2}+\frac12\expt{v \vb}\expt{\vb}
		- \frac12\expt{v}\expt{\vb^2} -\frac14\expt{v}\expt{\vb}^2
		+\frac34\expt{\vb}\pd^a\l^2\pd_a\bar\l^2 \big) ~.
\end{align}

\section{Deriving the nonlinear field redefinition}\label{sect:AVtoNL}

In this section, we sketch the derivation of (\ref{eqn:AVtoNL_simp}).
A more detailed presentation of our method will be given in a separate publication.

Our goal is  to find a nonlinear field redefinition 
$\l_\a \to \l_\a(\j,\bar\j) = \j_\a + O(\k^2)$
that satisfies
\begin{align} 
	S_\AV[\l(\j,\bar\j),\bar\l(\j,\bar\j)] \equiv \tilde S_\AV[\j,\bar\j]
	= S_\KS[\j,\bar\j]~.
\end{align}
Since both actions $S_\AV[\l,\bar\l]$ and $S_\KS[\j,\bar\j]$
are invariant under $R$-symmetry,
the nonlinear transformation we are looking for must be 
covariant under $R$-symmetry.
The most general field transformation of this type is 
\begin{align} \label{eqn:GeneralFieldRedef}
	\l_\a (\j,\bar\j) 
	&= \j_\a+{\k^2}\j_\a\expt{\a_1 u + \a_2\ub} 
		+\rmi{\k^2}(\s^a\bar\j)_\a(\pd_a\j^2)\Big( \a_3
		+{\k^2}\expt{\b_7 u + \b_8 \ub}	\Big)  \\\non
	&+{\k^4}\j_\a\big(\b_1\expt{u\ub}+\b_2\expt{u}\expt{\ub}
		+ \b_3\expt{\ub^2} + \b_4 \expt{\ub}^2
		+ \b_5\pd^a\j^2\pd_a\bar\j^2 + \b_6\bar\j^2\Box\j^2\big) \\\non
	&+ {\k^6}\j_\a\big(\g_1\expt{u\ub^2}+\g_2\expt{u \ub}\expt{\ub}
		+ \g_3\expt{u}\expt{\ub^2} + \g_4\expt{u}\expt{\ub}^2
		+\g_5\expt{\ub}\pd^a\j^2\pd_a\bar\j^2 \big) ~.
\end{align}
This is equivalent to the field redefinition used in \cite{Kuzenko2005e} up to 
some 7-fermion identities.

The general field redefinition at $O(\k^2)$
acts on the AV action to give
\begin{align} 
\begin{aligned}
	\tilde S_\AV
	&= -\intx\Big\{\frac12\expt{u+\ub} 
	+{\k^2}\Big(\frac12\big(\expt{u}^2-\expt{u^2}+\cc\big)
	+\big((\a_1+\a_3)\expt{u}^2 + \cc\big) \\
	&\quad-	\big(\a_3\expt{u^2} + \cc\big) 
	+ 2\Re(\a_2)\expt{u}\expt{\ub}-\Re(\a_3)\pd^a\j^2\pd_a\bar\j^2\Big)
	+ O(\k^4)\Big\} \ ,
\end{aligned}
\end{align}
where we have rewritten all terms in the minimal basis
\begin{align} \label{eqn:4fermion_basis}
	\expt{u^2},	\quad	\expt{\ub^2},	\quad 	\expt{u}\expt{\ub}, \quad
	\expt{u}^2,	\quad	\expt{\ub}^2,	\quad	\pd^a\j^2\pd_a\bar\j^2~.
\end{align}
Obviously, if we are to match $S_\KS$ to this order we need
\begin{align} \label{eqn:Match_at_Ok2}
	\qquad	\a_1&=0~,& 			\Re(\a_2)&=0~,&
			\a_3&=-\frac12~,& 	2f^2&=\k^{-2}~.\qquad
\end{align}
The imaginary part of $\a_2$,
which we will denote as $\a_2^\rmi$, 
is not fixed at this order.

The effect of \eqref{eqn:GeneralFieldRedef} with \eqref{eqn:Match_at_Ok2}
on the AV action at $O(\k^4)$ can be similarly analysed.
If we split all coefficients into their
real and imaginary parts, $\b_j = \b_j^\rmr + \rmi \b_j^\rmi$, 
then the restrictions on the $\b_j$ can be written as
\begin{gather} \label{eqn:Match_at_Ok4}
	\b_1^\rmr = 4\b_6^\rmr + 2\b_8^\rmr~,\quad 
	\b_1^\rmi = 2\a_2^\rmi + 4\b_6^\rmi - 2\b_8^\rmi~,\quad
	\b_3^\rmr = -\frac1{2}\big(1+4\b_6^\rmr\big)~, \quad
	\b_3^\rmi = -2(\a_2^\rmi + \b_6^\rmi)~,  \non \\
	\b_2^\rmr = \frac32 - \b_4^\rmr + 4\b_6^\rmr - \b_7^\rmr - \b_8^\rmr~,\quad
	\b_2^\rmi = -\frac12\a_2^\rmi + \b_4^\rmi -\b_7^\rmi + \b_8^\rmi~, \\\non
	\b_5^\rmr = \frac12+2\b_6^\rmr + \b_8^\rmr~, \quad
	\b_5^\rmi = \a_2^\rmi + 2\b_6^\rmi - \b_8^\rmi~,\quad
	\b_6^\rmr = -\frac1{8}\big(1+(\a_2^\rmi)^2\big)~.
\end{gather}
The seven real parameters
$\b_4^\rmr$, $\b_4^\rmi$, $\b_6^\rmi$, $\b_7^\rmr$, 
$\b_7^\rmi$, $\b_8^\rmr$ and $\b_8^\rmi$
are not fixed at this order.

A similar analysis is performed at $O(\k^6)$ and we find
that to match  $\tilde S_\AV$ to $S_\KS$  we need
\begin{equation}\label{eqn:Match_at_Ok6}\begin{aligned}
	\g_1   &= 1~,\quad
	\g_2^\rmr = \tfrac3{2}
			-2\a_2^\rmi\big(\tfrac14\a_2^\rmi+2\b_6^\rmi-\b_8^\rmi\big)
			+ 2(\b_7^\rmr + \b_8^\rmr + \g_5^\rmr)~,\\
	\g_3^\rmr &= \tfrac{3}{4} 
			-\a_2^\rmi\big(\tfrac14\a_2^\rmi+2\b_6^\rmi+\b_7^\rmi-\b_8^\rmi\big)
			+ 2\b_4^\rmr + 3\b_8^\rmr~,\\	
	\g_3^\rmi &= -\a_2^\rmi\big(\tfrac14(\a_2^\rmi)^2
				+\tfrac9{4}-\b_7^\rmr-\b_8^\rmr\big)
			-2 \b_4^\rmi - 6\b_6^\rmi - \b_8^\rmi~,\\
	\g_4^\rmr &= -\a_2^\rmi\big(\a_2^\rmi+\tfrac32\b_4^\rmi+2\b_6^\rmi
			+\tfrac12\b_7^\rmi + \tfrac32\b_8^\rmi\big) 
			+ \tfrac12\big(\b_4^\rmr - \b_7^\rmr + \b_8^\rmr\big)~.
\end{aligned}\end{equation}
The free parameters at this order are 
$\g_2^\rmi$, $\g_4^\rmi$, $\g_5^\rmi$ and $\g_5^\rmr$,
the first three of which have completely dropped out the calculation. 

It can be shown that all the free parameters can be accounted
for by the symmetries of either one of the two actions. 
In particular $\g_2^\rmi$, $\g_4^\rmi$ and $\g_5^\rmi$ 
correspond to single term trivial
symmetries of any Goldstino action.

From the above results we see that out of the original 32 real parameters
in the nonlinear field redefinition, 12 remain unfixed by the requirement that
\( \tilde S_\AV = S_\KS\). Since these freedoms may be recovered
by a symmetry transformation of either action, we may simply set all 
free parameters to zero and get the field redefinition \eqref{eqn:AVtoNL_simp}.

Some results of this section were obtained with computer assistance \cite{mma}.
The core of the computer program is the generation of a 
canonical form for expressions involving spinors, 
which is necessary for comparing expressions. 
All Fierz-type identities were automatically satisfied 
by choosing a representation for the Pauli matrices 
and defining a definite ordering for spinors and their derivatives.
Total derivatives, where relevant, were removed from expressions by generating 
a set of replacement rules that performed the appropriate integration by
parts to yield a unique form for the expression.
Further details of the algorithm will be given in a separate publication.

\section{Concluding comments}
It has been pointed out, e.g. \cite{Casalbuoni1989,Liu2010}, 
that $S_\KS$ does not have definite transformation properties under
the supersymmetry transformation \eqref{eqn:Nonlin_SUSY}.  
But now that we have an explicit mapping from $S_\AV$ to $S_\KS$ 
we can use it to find the supersymmetry transformation under which 
$S_\KS$ is invariant. We get
\begin{align} \label{KS:SUSY}
	\d_\x \j_\a &= \d_\x \j_\a(\l,\bar\l) 
		= \d_\x\l^\b\cdot\frac{\d}{\d\l^\b}\j_\a(\l,\bar\l)
		+ \d_\x\bar\l_\db\cdot\frac{\d}{\d\bar\l_\db}\j_\a(\l,\bar\l)
			\Big|_{\l=\l(\j,\bar\j)} 	\\\non
		&= \frac1\k\x_\a - \rmi \k \left(
		\big(\j\s^a\bar\x-\x\s^a\bar\j\big)\pd_a\j_\a
		-(\s^a\bar\j)_\a\pd_a(\x\j) - \frac12(\s^a\bar\x)_\a\pd_a\j^2
		\right) + O(\k^3)~.
\end{align}

Finally, we would like to comment on the field redefinition found by 
Zheltukhin \cite{Zheltukhin2010,Zheltukhin:2010xr}. 
In these papers, written in the four-component spinor notation,
the required field redefinition was sought in the form%
\footnote{We use bold-face Greek letters for four-component spinors.}
${\bm \j}( {\bm \l})={\bm \l}+\k^2{\bm \c}({\bm \l})+O(\k^4)$. 
By requiring that $S_\KS[{\bm \j}( {\bm \l })]=S_\AV[{\bm \l}]$ 
a solution was found for $\bm \c$.
The key step in the $O(\k^2)$ calculation reported in \cite{Zheltukhin2010}
is the factorisation 
\begin{align} \label{Z:factorised}
	(\pd^m\bar{\bm \l})\big(\g_m {\bm \c}+{\bm \z}_m({\bm \l})\big) = 0 \,, 
\end{align}
where we have introduced
\begin{align}
{\bm \z}_m({\bm \l}) 
			= \frac\rmi2\big({\bm \l}(\bar{\bm \l}_{,m}{\bm \l})
			+ \g_5{\bm \l}(\bar{\bm \l}_{,m}\g_5{\bm \l})\big)
			-\frac\rmi{4}\big(\g_m{\bm \l}(\bar{\bm \l}_{,n}\g^n{\bm \l})
			-\g^n{\bm \l}(\bar{\bm \l}_{,n}\g_m{\bm \l})\big) 
\end{align}
and denoted by $\bar{\bm \l}_{,m} $ the derivative of 
$\bar{\bm \l}$ with respect to $x^m$.
In \cite{Zheltukhin2010,Zheltukhin:2010xr}, 
it was then inferred from \eqref{Z:factorised} that
\begin{align} \label{Z:inconsistent}
	\g_m {\bm \c}  + {\bm \z}_m({\bm \l}) = 0 \ . 
\end{align}
Unfortunately, the 16 equations  \eqref{Z:inconsistent} 
for the 4 components of $\bm \c$ are inconsistent.
This can be seen by taking a time- or space-like vector $p^m$ 
and contracting both sides of \eqref{Z:inconsistent} with 
$(p^n\g_n)^{-1}p^m =-p^{-2} (p^n\g_n) p^m$.
Then, the first term in the relation obtained will be $p$-independent, 
while the second remains $p$-dependent.

\vspace{2ex}
\noindent
{\bf Acknowledgements:}\\
SJT is grateful to Ian McArthur, Paul Abbott and Joseph Novak 
for useful discussions.
SMK acknowledges email correspondence with Alexander Zheltukhin. 
The work of SMK is supported in part by the Australian Research Council.


\providecommand{\href}[2]{#2}
\begingroup\raggedright
\endgroup

\end{document}